\documentclass[aps, prl, twocolumn, superscriptaddress]{revtex4}

\usepackage{graphicx} 
\usepackage{dcolumn} 
\usepackage{bm} 
\usepackage{amsmath}
\usepackage{SIunits}
\usepackage{times}
\usepackage[ansinew]{inputenc}
\usepackage{upgreek}
\usepackage{physics}
\usepackage{hyperref}

\begin{document}

\title{Observation of backscattering induced by magnetism in a topological edge state}

\author{Berthold J\"ack}
\affiliation{Princeton University, Joseph Henry Laboratory at the Department of Physics, Princeton, NJ 08544, USA}
\author{Yonglong Xie}
\affiliation{Princeton University, Joseph Henry Laboratory at the Department of Physics, Princeton, NJ 08544, USA}
\author{B. Andrei Bernevig}
\affiliation{Princeton University, Joseph Henry Laboratory at the Department of Physics, Princeton, NJ 08544, USA}
\author{Ali Yazdani}
\email[electronic address:\ ]{yazdani@princeton.edu}
\affiliation{Princeton University, Joseph Henry Laboratory at the Department of Physics, Princeton, NJ 08544, USA}

\date{\today}

\begin{abstract}
The boundary modes of topological insulators are protected by the symmetries of the non-trivial bulk electronic states. Unless these symmetries are broken, they can give rise to novel phenomena, such as the quantum spin Hall effect in one-dimensional (1D) topological edge states, where quasiparticle backscattering is suppressed by time-reversal symmetry (TRS). Here, we investigate the properties of the 1D topological edge state of bismuth in the absence of TRS, where backscattering is predicted to occur. Using spectroscopic imaging and spin-polarized measurements with a scanning tunneling microscope, we compared quasiparticle interference (QPI) occurring in the edge state of a pristine bismuth bilayer with that occurring in the edge state of a bilayer, which is terminated by ferromagnetic iron clusters that break TRS. Our experiments on the decorated bilayer edge reveal an additional QPI branch, which can be associated with spin-flip scattering across the Brioullin zone center between time-reversal band partners. The observed QPI characteristics exactly match with theoretical expectations for a topological edge state, having one Kramer's pair of bands. Together, our results provide further evidence for the non-trivial nature of bismuth, and, in particular, demonstrate backscattering inside a helical topological edge state induced by broken TRS through local magnetism.
\end{abstract}

\maketitle

\textbf{Introduction.}--Since the discovery of a  topological insulator (TI) state in two-dimensional (2D) mercury-telluride quantum wells more than a decade ago \cite{BAB_2006_1, Koenig_2007}, scientific interest and technological perspectives fueled research efforts on this topic, which lead to the discovery of three-dimensional topological insulators \cite{Fu_2007, Hsieh_2008}, and more recently, to the identification of crystalline and higher order TIs (HOTI), in which crystal symmetries support protected surface, edge and corner states \cite{Fu_2011, WAB_2017_1, WAB_2017_2, Langbehn_2017, Song_2017}.

\begin{figure}[hb!]
\centering
\includegraphics[height=11.4cm, width=8.6cm]{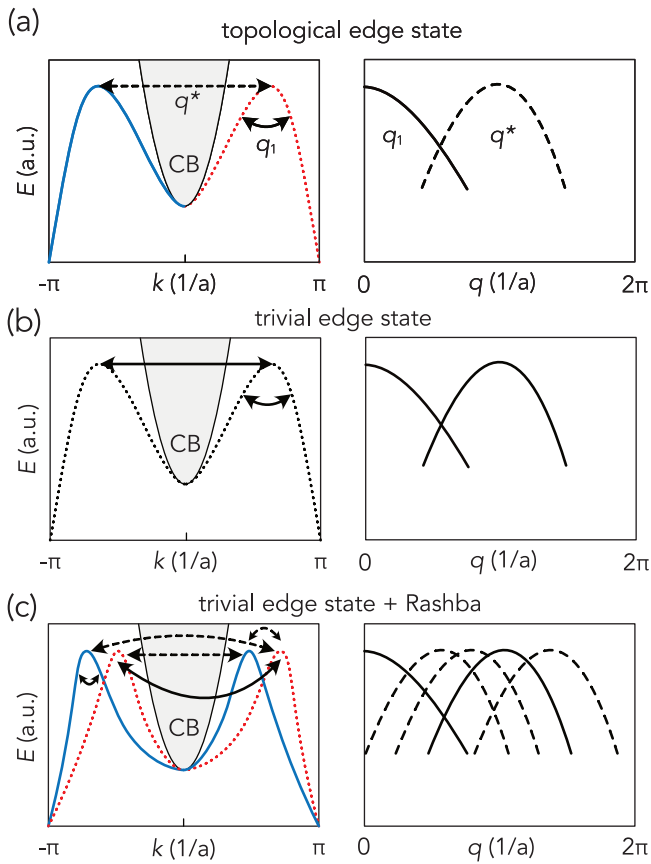}
\caption{(a) Left: Sketch of a 1D topological band structure near the band maximum (spin up -- red, dotted; spin down -- blue solid, CB - conduction band). Right: Sketch of theoretical QPI branches for spin-conserving $q_1$ (solid) and spin-flip $q^*$ (dashed) scattering, respectively. (b) Left: Sketch of a 1D trivial doubly spin degenerate band structure (dotted line) near the band maximum. Right: Sketch of theoretical QPI branches. (c) Left: Sketch of a trivial SOC split band structure near the band maximum (spin up -- red, dotted; spin down -- blue solid). Right: Sketch of theoretical QPI branches for spin-conserving scattering (solid) and spin-flip scattering (dashed).}
\label{fig1}
\end{figure}

A key property of all classes of TIs is the existence of a topological boundary mode, which is protected by the symmetries of the non-trivial bulk. Unless these symmetries are broken, they support protected band crossings at the high symmetry points of the band structure. For example, the one-dimensional (1D) edge state of 2D TIs is protected by time-reversal symmetry (TRS), which suppresses backscattering and gives rise to the quantum spin Hall (QSH) effect \cite{BAB_2006_2, Roth_2009}. Conversely, breaking TRS lifts these protected band crossings, and suppresses the QSH effect as recent experiments on monolayer WTe$_2$ suggest \cite{Qiang_2014, Wu_2018}.

The suppression of backscattering between orthogonal spin states of topological boundary modes was also addressed in scanning tunneling microscopy (STM) experiments on the surface of 3D TIs \cite{Roushan_2009, Zhang_2009, Beidenkopf_2011, Alpichshev_2012}. More recent STM and quantum transport experiments on the edge state of bismuth (Bi) bilayers and nanowires reported the absence of backscattering and the observation of ballistic transport, consistent with its topological origin \cite{Murakami_2006, Drozdov_2014, Murani_2017}. Additional theoretical investigations identified bulk Bi as a HOTI, whose 1D topological edge modes run along the hinges of its 3D bulk \cite{Schindler_2018}, while others argue that bulk Bi is in the critical region between a weak and strong TI \cite{Nayak_2019}. 

Nevertheless, despite all efforts, direct experimental evidence of backscattering in STM experiments, such as induced by scattering from magnetic impurities that break TRS \cite{Qi_2008, Liu_2009}, remained elusive so far. For 2D topological surface states such demonstration is challenging as backscattering would occur near the Dirac point, where the Fermi surface is circular. In this case, backscattering from $k$ to $-k$ on this Fermi surface is near other scattering processes that are allowed, even if TRS is preserved \cite{Beidenkopf_2011}. One-dimensional topological edge modes, are therefore an ideal testbed for testing the role of TRS breaking in backscattering. 

In this letter, we investigate backscattering in the 1D topological edge state of Bi bilayers induced by ferromagnetic Fe clusters using quasiparticle interference (QPI) experiments with an STM. Bi bilayers are especially suited for such an experiment as the edge state band structure features pockets at finite momentum, favoring the observation of QPI even in the presence of TRS \cite{Roushan_2009, Beidenkopf_2011, Drozdov_2014}. Comparing our experimental QPI data with expected QPI patterns of trivial and topological edge states, we find direct evidence for backscattering induced by magnetism in the edge states of Bi.

\textbf{Theory.}--QPI experiments with an STM are a valuable tool to investigate the electronic band structure of materials \cite{Yazdani_2016}. Quasiparticle scattering from atomic step edges or impurities leads to interference patterns in real space, whose fast Fourier transform (FFT) can be related to scattering vectors within the band structure in reciprocal space. These scattering events are spin-sensitive in that scattering from a scalar potential, e.g. that created by an atomic step edge on the sample surface, only leads to scattering between states of same spin (see Fig.\,1(a)). If spin is transferred during the scattering process, e.g. by scattering from a magnetic cluster on the surface, additional scattering vectors are allowed that connect states of opposite spin (see Fig.\,\ref{fig1}(a)). This spin-selectivity of the QPI process can be used to distinguish topological from trivial edge states (as it was done in 2D by Roushan {\em et al.} in Ref.\,\cite{Roushan_2009}).

\begin{figure*}
\includegraphics[height=7.0cm, width=17.2cm]{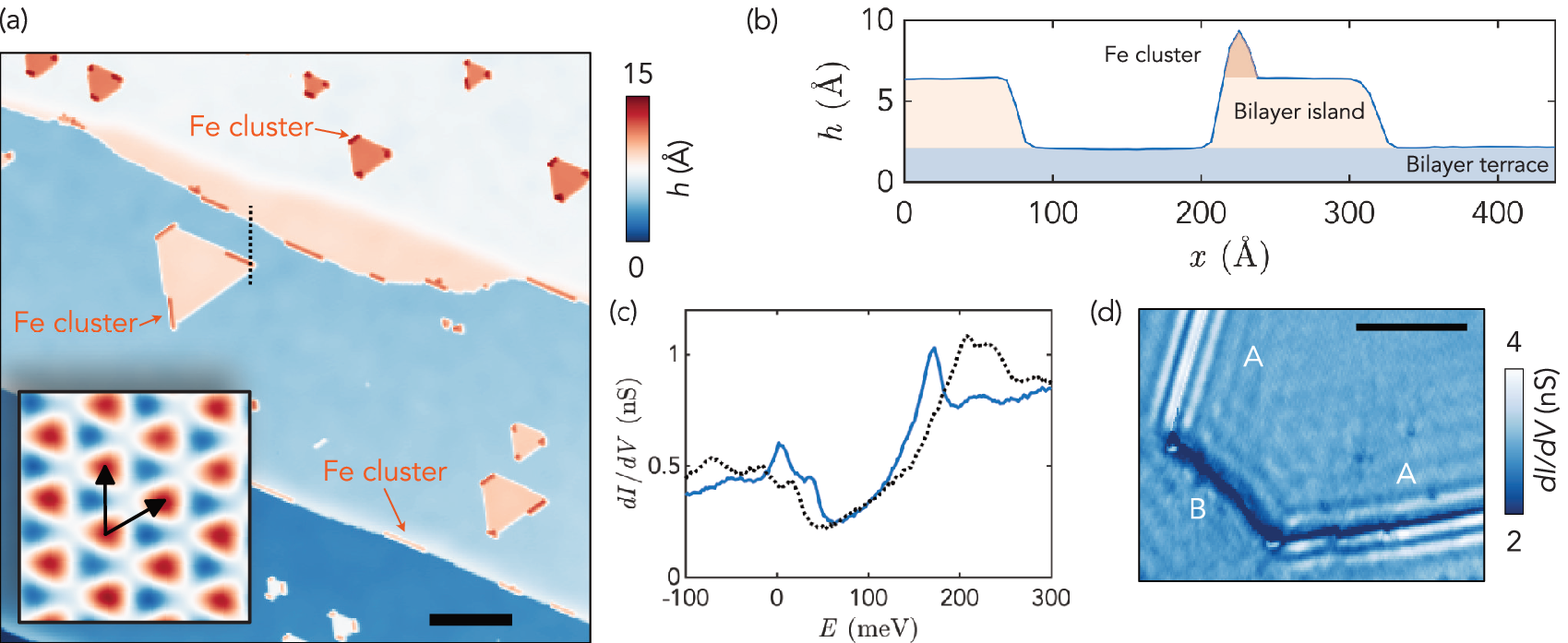}
\caption{(a) STM topography of the Bi(111) thin film surface measured under constant current condition ($V_{\textrm set}=-1\,$V, $I_{\textrm set}=50\,$pA, scale bar: 15 nm). Inset: Atomic resolution topography of a Bi(111) terrace measured under constant current condition ($V_{\textrm set}=10\,$mV, $I_{\textrm set}=1\,$nA). The two in-plane hexagonal atomic lattice vectors of Bi(111) are drawn as arrows. (b) Topographic line-cut along the black dashed line in (a). (c) $dI/dV$ point spectra measured with the STM tip located on a bilayer terrace (black, dashed line) and edge (blue, solid line), respectively ($V_{\textrm set}=300\,$mV, $I_{\textrm set}=750\,$pA, $V_{\textrm mod}=3\,$mV). (d) Constant current $dI/dV$ map with the bias fixed to $V_{\textrm set}=178\,$meV ($I_{\textrm set}=1\,$nA, $V_{\textrm mod}=3\,$mV, scale bar 25\,nm).}
\label{fig2}
\end{figure*}

A 1D topological state is characterized by one Kramer's pair of bands, which connects the bulk conduction and valence bands \cite{Hasan_2010}. Fig.\,\ref{fig1}(a) schematically illustrates a topological edge state band near its energetic maximum, which -- pertaining to the experimental results presented below -- was chosen to be similar to that of a Bi bilayer \cite{Drozdov_2014, LA_1995}. Spin-conserving intra-pocket scattering in this band results in a sole QPI branch emanating from $q=0$ (Fig.\,\ref{fig1}(a), right panel). If spin-flip scattering between the time-reversal partners is permitted, an additional, parabola shaped QPI branch emerges that is centered at finite scattering momentum $q\neq0$. 

A trivial 1D band is, in contrast, characterized by two Kramer's pairs owing to the spin degeneracy  (Fig.\,\ref{fig1}(b)). Therefore, both spin-conserving and spin-flip scattering results in two scattering vectors and cannot qualitatively be distinguished from another. Additional strong Rashba spin-orbit coupling (SOC), such as found on the Bi(111) surface \cite{Hofmann_2006}, splits these bands in momentum space (Fig.\,\ref{fig1}(c), left panel). Spin-conserving intra-pocket and inter-pocket scattering yield two scattering vectors, one with an onset at $q=0$ and another centered at $q\neq0$ (Fig.\,\ref{fig1}(c), right panel). If spin-flip scattering is permitted, three additional interband scattering vectors between bands of opposite spin momentum emerge, resulting in five different QPI branches. Hence, topological and trivial edge states can be distinguished by comparing the QPI induced by non-magnetic and magnetic scatterers, respectively.

\textbf{Results.}--Fig.\,\ref{fig2}(a) displays an STM topography of a sample surface typical for these experiments. It features uniform, defect-free terraces and trigonal bilayer islands, which exhibit a single bilayer step height of $\Delta z\approx 4.3\,\AA$ consistent with previous STM experiments on Bi(111) bulk crystals \cite{Drozdov_2014}. The atomic resolution topography (inset Fig.\,\ref{fig2}(a)) reveals the hexagonal arrangement of the Bi atoms in the bilayer, confirming the (111) direction of our epitaxially grown Bi film \cite{Hofmann_2006}. The deposition of Fe onto the surface leads to the decoration of the step edges with Fe clusters and chains, which are typically between 2 and 3 $\AA$ high, as measured from the top bilayer, and vary in their length along the island between a few to tens of nanometers (Fig.\,\ref{fig2}\,(b)).

\begin{figure}[ht!]
\centering
\includegraphics[height=6.5cm, width=8.6cm]{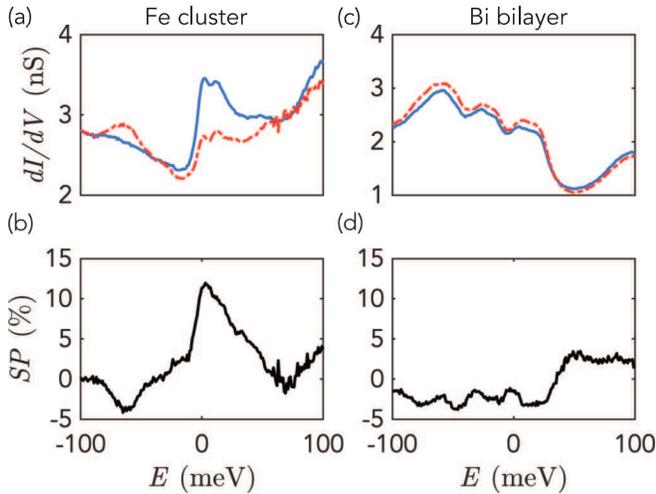}
\caption{$dI/dV$ spectra measured with 'UP' (solid line) and 'DOWN'(dashed line) polarized STM tips on top of an Fe cluster, (a) and on top of a bilayer terrace, (c) ($V_{\textrm{set}}=-100\,$mV, $I_{\textrm{set}}=1\,$nA, $V_{\textrm{mod}}=3\,$mV). (b) and (d) Spin polarization ($SP$) of the Fe cluster and Bi bilayer surface bilayer calculated from the data in (a) and (c), respectively.}
\label{fig3}
\end{figure}

We use scanning tunneling spectroscopy (STS) to demonstrate the presence of 1D bilayer edge states in our sample. In Fig.\,\ref{fig2}(c) we display $dI/dV$ spectra measured in the center of a surface terrace and on the edge of a bilayer island. The spectrum measured at the center features a prominent peak at $E=230\,\textrm{meV}$, which corresponds to the van-Hove singularity (vHs) of the 2D surface state \cite{Drozdov_2014, Hofmann_2006}. By contrast, the edge spectrum exhibits a vHs located at $E=178\,\textrm{meV}$. Mapping out the $dI/dV$ amplitude at this energy, shown in Fig.\,\ref{fig2}(d), demonstrates the 1D nature of this state confined to the bilayer edge. These observations agree with previous STM experiments on the Bi(111) surface, in which the vHs occurring at $E=178\,\textrm{meV}$ was associated with the hole-like maximum of the topological edge state band \cite{Drozdov_2014} (Fig.\,\ref{fig1}(a), left panel). The appearance of the edge state on every other edge, labelled A edge, in our STM experiments (Fig.\,\ref{fig2}(d)) can be understood in terms of its hinge-type character or its strong hybridization with the bulk, which leaves the edge state submerged in the bulk along the B-type bilayer edge \cite{Drozdov_2014,Schindler_2018}.

We studied the magnetic properties of the Fe clusters adsorbed to the bilayer edges using spin-polarized measurements with a magnetic STM tip \cite{Roland_2009}. To this end, we recorded STS spectra at different spin polarizations of the tip, labelled 'UP' and 'DOWN', respectively which has been trained with large out-of-plane magnetic fields of $\pm$1\,T. The spectrum measured on top of a Fe cluster with an 'UP' polarized tip features a pronounced peak at small positive bias voltages ($V_{\textrm set}\leq20\,$mV) with a sharp flank towards negative energies (Fig,\,\ref{fig3}(a)). In comparison, the spectrum measured with a 'DOWN' polarized tip exhibits similar features, however, with the peak amplitude much reduced. 

\begin{figure}[ht!]
\centering
\includegraphics[height=14.2 cm, width=8.7cm]{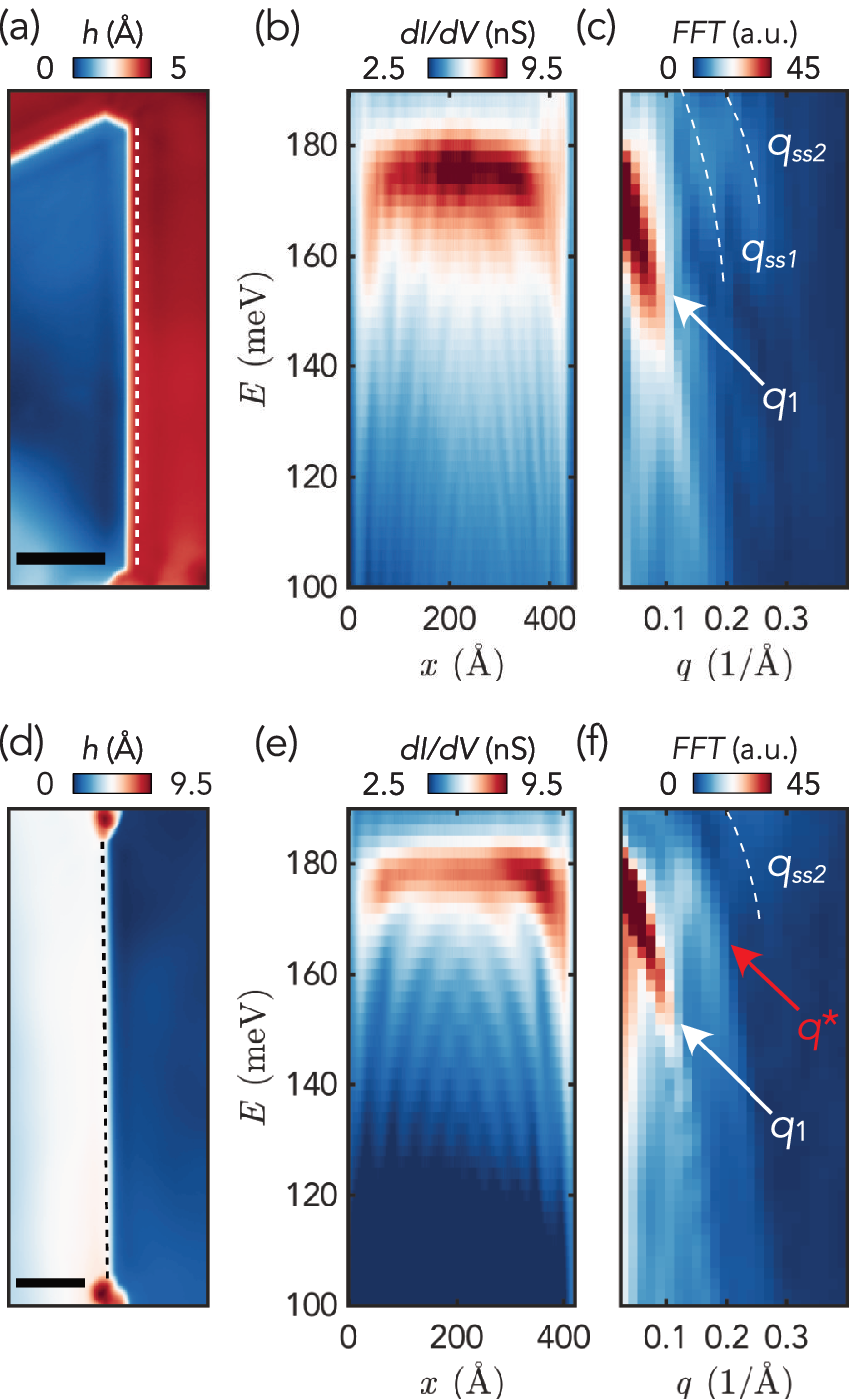}
\caption{(a) Surface topography of a pristine bilayer edge ($V_{\textrm set}=-1\,$V, $I_{\textrm set}=50\,$pA, scale bar 7\,nm). (b) Spectroscopic line cut taken along the white dashed line in (a) ($V_{\textrm set}=-100\,$mV, $I_{\textrm set}=1\,$nA, $V_{\textrm mod}=3\,$mV). (c) FFT of the real space data in (b). (d) Surface topography of a decorated bilayer edge ($V_{\textrm set}=-1\,$V, $I_{\textrm set}=50\,$pA, scale bar 5\,nm). (e) Spectroscopic line cut taken along the black dashed line in (d) ($V_{\textrm set}=-100\,$mV, $I_{\textrm set}=1\,$nA, $V_{\textrm mod}=3\,$mV). (f) FFT of the real space data in (e).}
\label{fig4}
\end{figure}

The calculated spin-polarization (SP) of that feature is $SP\approx12\,\%$ (Fig,\,\ref{fig3}(b)) and demonstrates the ferromagnetic nature of the Fe clusters, consistent with our recent work on this sample platform \cite{Jaeck_2019}. The pronounced LDOS peak with high SP can be attributed to the hybridization of the Fe $d$-orbitals with the underlying edge state near Fermi energy \cite{Jaeck_2019}. As a measure of consistency, spectra measured with 'UP' and 'DOWN' polarized tips located a few nanometers away from the cluster on top of the Bi bilayer terrace appear almost identical, except a marginal spin polarization of $SP<3\,\%$ (Fig,\,\ref{fig3}(c) and (d)). We have investigated another comparable Fe cluster using spin-polarized measurements, the results of which are consistent with the data presented in Fig.\,\ref{fig3} (please refer to Ref.\,\cite{SI} for the supplementary data set; a more detailed study of the Fe cluster magnetization properties can be found in Ref.\,\cite{Jaeck_2019}).

We used QPI measurements to investigate the topological character of the bilayer edge states. The necessary sample conditions and drift-stability requirements to carry out these demanding high-resolution measurements are described in detail in Ref.\,\cite{SI}. First, we investigate the QPI occurring in a pristine bilayer A edge prior to the Fe evaporation (Fig.\,\ref{fig4}(a)), in which case only spin-conserving QPI is expected to occur. We recorded $dI/dV$ spectra in an energy window around the edge state vHs (c.f. Fig.\,\ref{fig2}(c)) as a function of the tip position along the bilayer edge. The result of this spectroscopic line cut is displayed in Fig.\,\ref{fig4}(b). Starting from the vHs at $E=178\,$meV, the data feature a spatially oscillating QPI pattern along the bilayer edge and extending towards lower energies, reminiscent of a 1D particle in a box behavior. The FFT of the spectroscopic line cut is shown in Fig.\,\ref{fig4}(c). It is dominated by a sole QPI branch $q_1$, which originates at the vHs maximum, $q(178\,$meV$)=0\,\AA^{-1}$, and exhibits a negative slope. We also detect other features of much reduced amplitude, $q_{\rm ss1}$ and $q_{\rm ss2}$, which are weakly dispersing from $E>200\,$meV at $q\geq0.2\,\textrm{\AA}^{-1}$.

\begin{figure}[t]
\centering
\includegraphics[height=6.0cm, width=8.6cm]{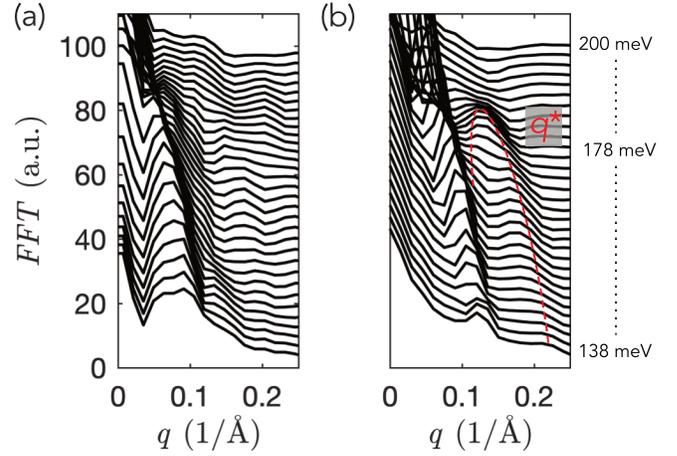}
\caption{(a) Waterfall plot (curves vertically offset by 1 a.u.) of the QPI on the pristine edge (Fig.\,\ref{fig4}(c)) in the energy window $[138...200]\,$meV. (b) Waterfall plot (curves vertically offset by 1 a.u.) of the QPI on the decorated edge (Fig.\,\ref{fig4}(f)) in the energy window $[138...200]\,$meV.}
\label{fig5}
\end{figure}

To probe possible spin-flip scattering events, we studied QPI in a bilayer edge terminated by ferromagnetic Fe clusters at both ends (Fig.\,\ref{fig4}(d)). The spectroscopic line cut along this edge (Fig.\,\ref{fig4}(e)) features a rich QPI pattern similar to that of the pristine edge, but with a more complex QPI pattern at energies below the vHs maximum. This becomes more evident in the FFT of these data presented in Fig.\,\ref{fig4}(f): In addition to $q_{1}$, also observed on the pristine edge, the decorated edge features a previously unseen, parabola shaped QPI branch $q^*$, which has its energy maximum at $q=0.15\pm0.02\,\textrm{\AA}^{-1}$ (see Fig.\,\ref{fig5}(b)). Similar to the QPI on the pristine bilayer edge, also the QPI on the decorated features a faint signal, $q_{\rm ss2}$, weakly dispersing from $E>200\,$meV at $q\geq0.2\,\textrm{\AA}^{-1}$ (Fig.\,\ref{fig4}(f)).

\textbf{Discussion.}--The reproducibility of our main experimental results presented in Fig.\,\ref{fig4} underlines their robust nature (please refer to Ref.\,\cite{SI} for the supplementary data sets). Comparing our observed QPI patterns with previous experiments on pristine Bi edge states \cite{Drozdov_2014}, we identify $q_1$ as resulting from scattering inside the hole pocket near the edge state band maximum at $E=178\,$meV (Fig.\,\ref{fig1}(a)). We also observe good agreement with Drozdov's {\em et al.} results in terms of the vanishing QPI amplitude towards lower energies for both type of edges (Fig.\,\ref{fig4}(c) and (f)), owing to the SOC induced spin texture \cite{Hofmann_2006}. The absence and presence of $q^*$ on pristine and decorated edges, respectively, its onset at the same energy maximum as $q_1$ and its centering at $q\neq0$ are highlighted by the waterfall plots in Fig.\,\ref{fig5}(a) and (b). Together these observations strongly indicate that $q^*$ originates from spin-flip scattering between TRS partners, induced by the ferromagnetic Fe clusters. We emphasize that $q^*$ cannot originate from scattering inside the hole-type pocket of the surface state band maximum, which was previously found to reside at $E>200$\,meV (cf.\,Fig.\,\ref{fig2}(c)) \cite{Hofmann_2006, Drozdov_2014, Du_2016}. By contrast,  we can associate the faint QPI branches $q_{\rm ss1}$ and $q_{\rm ss2}$ (Fig.\,\ref{fig4}(c) and (f)) with such surface state QPI, which projects onto the bilayer edge and alters the measured edge state $dI/dV$-spectra. A previous detailed analysis identified these features to result from scattering between the electron and hole pockets of the surface state band structure, residing at the $\Gamma$ and $M$ points, respectively \cite{Drozdov_2014}.

The interpretation of our observations becomes evident when comparing the experimental QPI features with theoretical expectations for different types of band structures in Fig.\,\ref{fig1}. The experimental QPI characteristic ($1\rightarrow2$ branches, pristine $\rightarrow$ decorated) matches the expectations for scattering between the TRS band partners of a topological edge state (Fig.\,\ref{fig1}(a)), and cannot be explained by scattering within a trivial edge state band structure neither with ($2\rightarrow5$, Fig.\,\ref{fig1}(c)) nor without ($2\rightarrow2$, Fig.\,\ref{fig1}(b)) Rashba SOC.

While previous results on the bilayer edge state, such as the absence of backscattering \cite{Drozdov_2014}, were consistent with the properties of a topological edge state, the observation of backscattering induced by magnetism together with the overall QPI characteristics in this study provide direct evidence for its non-trivial origin \cite{Murakami_2006, Schindler_2018, Nayak_2019}. We note that our experimental results demonstrate the existence of one Kramer's pair of bands on the edge of an individual Bi bilayer. However, our measurements cannot distinguish between a higher order and possible weak topological classification of the bulk that were debated recently \cite{WAB_2017_1, WAB_2017_2, Nayak_2019}.

Finally, we address the dependence of quasiparticle backscattering on the Fe cluster magnetization. Model calculations show that ferromagnetic clusters, which have any magnetization component perpendicular to the quasiparticle spin, will always induce backscattering in the helical topological edge mode \cite{SI}. Our previous detailed investigation of the magnetic cluster magnetization on this platform reveals their complex magnetization profile, whose vector is not aligned with the calculated edge state quasiparticle spin \cite{Jaeck_2019}. Hence, these Fe clusters promote the occurrence of backscattering, consistent with our experimental observations presented in Fig.\,\ref{fig4}(f).

\textbf{Conclusion.}--We investigated QPI in the edge states of Bi bilayers with a STM. QPI observed on pristine edges describes spin-conserving scattering inside a hole-type pocket of the edge state band. QPI in edge states of bilayers decorated with ferromagnetic Fe clusters features an additional QPI branch that can be associated with spin-flip scattering between time-reversal band partners. The experimental QPI characteristics match with theoretical expectations for a topological edge state having one Kramer's pair of bands, and provide direct evidence for backscattering in the topological edge states of Bi. Looking ahead, we envision experiments to study the dependence of backscattering on the magnetization direction of tailored ferromagnetic clusters \cite{SI}. Ultimately, such experiments could realize nanoscale magnetic switches to control charge carrier transport (ballistic vs. resistive) along the topological 1D channel towards future spintronic device applications. 

\textbf{Acknowledgements.}--The authors acknowledge helpful discussions with Raquel Queiroz, Jian Li and Sangjun Jeon. This work has been supported by NSF-DMR-1904442, ONR-N00014-17-1-2784, ONR-N00014-14-1-0330, NSF-MRSEC programs through the Princeton Center for Complex Materials DMR-142054, NSF-DMR-1608848, and the Gordon and Betty Moore Foundation as part of EPiQS initiative (GBMF4530). Additional support has come from the Alexander-von-Humboldt foundation (BJ) as well as from the DOE de-sc0016239, the NSF EAGER 1004957, a Simons Investigator Grant, ARO MURI W911NF- 12-1-0461, the Packard Foundation, and the Schmidt Fund for Innovative Research (BAB).

\textbf{Methods.}--We grew a nominal 40 $\AA$ thick Bi(111) film on a clean niobium(110) surface cooled to liquid nitrogen temperatures during deposition and subsequently post-annealed to $T = 373\,$K for $t = 20\,$min. Iron (Fe) clusters were synthesized by depositing $\leq0.5$ monolayer Fe on the prepared Bi surface at $T = 373\,$K. STM experiments were performed using a home-built ultra-high vacuum STM operating at an electron temperature of $T = 1.4\,\textrm{K}$. We used cut PtIr STM tips that were treated by field emission and conditioned on a Cu(111) surface. Bulk chromium tips for spin-polarized STM experiments were conditioned on Fe nanostructures grown on tungsten. The STM topographies are recorded at constant current conditions and differential tunnel conductance $dI/dV$ curves and spectroscopic line cuts at open feedback conditions, respectively using standard lock-in instrumentation (modulation voltage $V_{\textrm mod}$, frequency $f_{\textrm mod}=4\,\textrm{kHz}$, time constant $t_{\textrm mod} = 5\,\textrm{ms}$). Appropriate bias and current set points, $V_{\textrm{set}}$, and $I_{\textrm{set}}$, respectively were chosen for all measurements. The experimental data and the related analysis code, required to reproduce the content of this manuscript, are accessible through Ref.\,\cite{datarep}.

\bibliography{refs} 

\end{document}